\documentclass[showpacs,preprintnumbers,amsmath,amssymb,prl,twocolumn]{revtex4}

\usepackage{graphicx}
\usepackage{dcolumn}
\usepackage{bm}
\def\Nb{N_{}}
\def\Nf{N_{\rm F}}
\def\QED{\rm F}

\def\comment#1{}
\begin{document}

\title{
Critical behavior of Ginzburg-Landau model coupled to massless
Dirac fermions
}
\author{Hagen Kleinert}
\email{kleinert@physik.fu-berlin.de}
\homepage{http://www.physik.fu-berlin.de/~kleinert/}
\author{Flavio S. Nogueira}
\email{nogueira@physik.fu-berlin.de}
\affiliation{Institut f\"ur Theoretische Physik,
Freie Universit\"at Berlin, Arnimallee 14, D-14195 Berlin, Germany}

\date{Received \today}

\begin{abstract}
We point out interesting effects
of additional massless Dirac fermions with $\Nf$ colors upon
 the critical behavior of the Ginzburg-Landau model. For increasing
$\Nf$,
the
model is driven
into the
type II regime of superconductivity.
The critical exponents are given as a function of $\Nf{}$.
\end{abstract}


\maketitle

The critical fluctuations in the Ginzburg-Landau (GL) model of
superconductors are
an old problem in condensed matter physics \cite{HLM}.
While the underlying  complex order field
theory with $|\phi|^4$-interaction
is well understood, \cite{KS}
no satisfactory approximation has been found
for a long time to deal with
the additional gauge field.
This may seem surprising
since the Lagrangian is quadratic on the gauge field ${\bf A}$.
One has therefore
 expected
that ${\bf A}$
can be integrated out in a reasonable approximation
to obtain an effective action
with extra terms in the order field $\phi$. \cite{HLM}
This is exactly possible for constant $|\phi|$
where a
mean-field
approximation for the effective potential
receives an extra term
$\sim -|\phi|^d$ in $d$ dimensions. In four dimensions,
where the
model is relevant to particle physics, the extra term is
$\sim |\phi|^4\ln |\phi|^2$. \cite{CW} Such an extra term, if  present
in the full effective potential,
would  make the second-order phase
transition first-order.
A similar conclusion is derived from
a one-loop renormalization
group (RG) calculation in $d=4-\epsilon$ dimensions, which shows no
 non-trivial charged fixed point
\cite{HLM,Lawrie}, even up to two loops \cite{Tess,Koln}.
If the GL model is generalized in a such a way as to
contain $\Nb{}/2$ complex scalars instead of one, then non-trivial charged
fixed points are found at one loop for $\Nb{}>\Nb{}_c=365.9$.
Duality arguments, however, point out to the existence of a 
second-order phase transition at $N=2$ and in the type II regime 
\cite{Dasgupta,Kleinert}. 

A more significant reduction
of the critical value of $\Nb{}$ is achieved
by a RG approach in a fixed dimension $d\in(2,4)$.
As we shall see,
a one-loop
calculation in $d=3$ reduces
$\Nb{}_c$ less than a third of the above value.
This leads us to expect
that non-trivial
charged fixed points are more accessible in $d=3$
than in $4- \epsilon $ dimensions.
Indeed, this
was recently
confirmed  by the present authors
\cite{KleinNog}
by finding such a
fixed point at $\Nb{}=2$
 in a
new three-dimensional RG calculation {\em below\/} $T_c$.
The success of this approach relies on the explicit
presence of two mass scales in the problem, defined by the inverse of the
correlation length $\xi$ and penetration depth $\lambda$. This is in contrast
to all previous studies which were done
in the disordered phase
at $T\geq T_c$ which has only
one physical mass scale.
It has been proposed
to introduce a second scale by assuming different
renormalization points for each coupling of the theory \cite{Herbut}.
Such a procedure has in fact led
to charged fixed points at
$\Nb{}=2$,~$d=3$ provided the ratio between the two scales is
sufficiently large
 \cite{Herbut}. However,
 this ratio must be
fixed by conditions external to the formalism, and a
duality result
\cite{Kleinert} for a tricritical point was used to do so.
Thus, in contrast to the recent
finding for $T< T_c$,
the RG-theoretic
 explanation of a
charged fixed point for $T\geq T_c$ at $\Nb{}=2$
has remained
obscure.

The purpose of this paper is  to
point out the interesting modifications of the
critical behavior of the GL model with $N/2$ complex field
$\phi$ brought about by the presence of
extra massless Dirac fermion fields $\psi$ with $\Nf{}$ replica.
The introduction of massles Dirac fermions is of great interest,
since effective microscopic models of strongly correlated
electrons usualy contain them \cite{Marston,Kim}.

The bare Lagrangian is assumed to be

\begin{equation}
\vspace{-5mm}
\label{L}
{\cal L}={\cal L}_{\rm \QED{}}+{\cal L}_{\rm GL},
\vspace{-2mm}
\end{equation}

\begin{equation}
\label{QED}
\vspace{-5mm}
{\cal L}_{\rm \QED{}}=\bar{\psi}_0\gamma_\mu(\partial_\mu+ie_0 A_\mu^0)\psi_0,
\vspace{-2mm}
\end{equation}

\begin{equation}
\label{GL}
{\cal L}_{\rm GL}=\frac{1}{4}F_{\mu\nu}^2+|(\partial_\mu-ie_0 A_\mu^0)\phi_0|^2
+m_0^2|\phi_0|^2+\frac{u_0}{2}|\phi_0|^4,
\end{equation}
where the subscript zero denotes bare quantities and
$F_{\mu\nu}=\partial_\mu A_\nu^0-\partial_\nu A_\mu^0$.
The labels for the $\Nf$ fermion and $\Nb$ boson replica are omitted.

The fermions have the effect of modifying
the gauge field properties of the GL model by giving it an
effective
non-local
gradient energy. Indeed, integrating out the fermions
generates a leading long-wavelength
energy

\begin{equation}
\label{nl}
{\cal L}_{\rm
eff}=\frac{\Nf{}}{16}F_{\mu\nu}\frac{1}{\sqrt{-\partial^2}}F_{\mu\nu}.
\end{equation}
In the infrared, this leading term makes the
initial Maxwell term in (\ref{GL}) irrelevant.
Since (\ref{nl}) gives the gauge field a unit dimension instead
of the canonical 1/2,
 the
 charge becomes effectively dimensionless.
Thus,  by integrating out the gauge field for a uniform order
field $\phi_0=\bar{\phi}$, we obtain the effective potential:

\begin{eqnarray}
\label{effpot}
V_{\rm eff}&=&\left(m_0^2+\frac{2e_0^2\Lambda^2}{3\pi^2\Nf{}}\right)
|\bar{\phi}|^2+\left(\frac{u_0}{2}-\frac{32e_0^4\Lambda}{
3\pi^2\Nf{}^3}\right)|\bar{\phi}|^4\nonumber \\&&-
\frac{256e_0^6}{3\pi^2\Nf{}^3}|\bar{\phi}|^6\ln
\left(\frac{8e_0^2|\bar{\phi}|^2}{\Nf{}\Lambda}\right),
\end{eqnarray}
where $\Lambda$ is an ultraviolet cutoff.
Note the important difference
with respect to the effective potential
of the usual GL model, where a term
$|\phi|^3$ is generated in three dimensions \cite{HLM},
giving rise to an apparent first-order transition.
The limit $\Nf\rightarrow 0$ is singular in this approximation,
which ignores  the Maxwell term
 controlling the gauge field fluctuations
for $\Nf=0$.

For large $\Nf{}$, $V_{\rm eff}$
reduces to the mean-field effective potential
of the pure $|\phi|^4$ theory. This should be not 
surprising since decoupling
by rescaling the charge $e_0\to e_0/\sqrt{\Nf{}}$ and taking the 
large $\Nf{}$ limit leads to an extreme type II superconductor
coinciding with the $O(\Nb{})$ model. Thus, $\Nf$ Dirac fermions
allows a novel
interpolation between the usual GL model and the
$O(\Nb{})$ model
which runs through
different intermediate physical
systems than the simple limit $e^2_0\rightarrow 0$.
It is therefore an interesting problem to study their
effect upon the critical behavior
as the number $\Nf{}$ is varied
for fixed $\Nb{}$.
This is what will be done
in this paper using
 RG techniques. At one loop and in
$d=4-\epsilon$ dimensions we find that
for $\Nb{}=2$, which is the physical number for a superconductor,
an infrared stable charged fixed point exists for $\Nf{}>\Nf{}_c=3.47$.
We repeat the
 study
 at fixed dimensions
$d\in(2,4)$ where we find that
 the critical number of fermions for $d=3$  is almost the same: $\Nf{}_c=4.44$
such that we can give the scheme-independent estimate
$\Nf{}_c=4\pm0.5$. Finally, all independent
critical exponents will be listed
as a function of $\Nf{}$.

Taking into account the Ward identities
due to
gauge invariance, the Lagrangians (\ref{QED}) and
(\ref{GL}) can be written in terms of renormalized quantities as

\begin{equation}
\vspace{-2mm}
\label{QEDR}
{\cal L}_{\rm \QED{}}=Z_\psi\bar{\psi}\gamma_\mu(\partial_\mu+ie A_\mu)\psi,
\vspace{-5mm}
\end{equation}

\begin{equation}
\vspace{-0mm}
\label{GLR}
{\cal L}_{\rm GL}=\frac{Z_A}{4}F_{\mu\nu}^2+
Z_\phi|(\partial_\mu-ie A_\mu)\phi|^2
+Z_m m^2|\phi|^2+\frac{Z_u u}{2}|\phi|^4.
\end{equation}
We define the dimensionless couplings 
$f\equiv \mu^{-\epsilon}Z_A e_0^2$ and $g\equiv \mu^{-\epsilon}Z_\phi^2
Z_u^{-1} u_0$, 
and fix the renormalization constants by minimal
subtraction of $1/ \epsilon^n $ pole terms.
The resulting one-loop $\beta$-functions
are

\begin{equation}
\vspace{-2mm}
\label{betaf1}
\beta_f=-\epsilon f+\frac{8\Nf{}+\Nb{}}{48\pi^2}f^2,
\end{equation}
\begin{equation}
\label{betag1}
\beta_g=-\epsilon g-\frac{3fg}{4\pi^2}+\frac{\Nb{}+8}{16\pi^2}g^2+
\frac{3}{4\pi^2}f^2.
\end{equation}
For $\Nf{}=0$ we recover the usual one-loop $\beta$-functions
of the GL model \cite{HLM,Lawrie}. Note that $\beta_g$
is unaffected by the fermions, being just the one-loop $\beta$-function
of the GL model.

The fixed points lie at

\begin{equation}
\vspace{-2mm}
f^*=\frac{48\pi^2\epsilon}{8\Nf{}+\Nb{}},
\end{equation}

\begin{figure}
\centering
{\includegraphics[width=7cm,angle=-90]{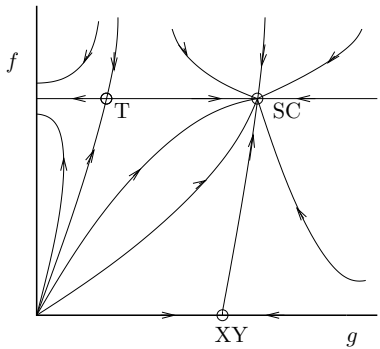} }
\vspace{-1.5cm}
\caption{Schematic flow diagram for the cases where the
$\Nf{}$ and $\Nb{}$ values are such that charged fixed points are generated.}
\label{Flow}
\end{figure}

\begin{widetext}
\begin{equation}
\label{gfix}
g^*_\pm=\frac{\epsilon\pi^2}{2}
\frac{576\Nb{}+16\Nb{}^2+4608\Nf{}+256\Nb{}\Nf{}+1024\Nf{}^2\pm
16(8\Nf{}+\Nb{})\sqrt{\Delta}}{
8\Nb{}^2+\Nb{}^3+128\Nb{}\Nf{}+16\Nb{}^2\Nf{}+512\Nf{}^{2}+64\Nb{}\Nf{}^{2}},
\end{equation}
\end{widetext}
where

\begin{equation}
\label{discr}
\Delta=-2160-360\Nb{}+\Nb{}^2+576\Nf{}+16\Nb{}\Nf{}+64\Nf{}^{2}.
\end{equation}
Accessible charged fixed points are obtained only if $\Delta>0$. The
case of interest for superconductivity is $\Nb{}=2$
for which Eq. (\ref{discr}) gives, under the
condition $\Delta>0$, a  charged fixed point
if

\begin{equation}
\label{Nc}
\Nf{}>{\Nf}_c=\frac{6\sqrt{30}-17}{4}\approx 3.47.
\end{equation}
We see that the number of fermions does not need to be large in
order to produce charged fixed points. A schematic flow diagram
is shown in Fig. \ref{Flow}. It exhibits precisely the
fixed point structure expected for the GL model
\cite{Lawrie,Herbut,Folk}. It has an infrared stable fixed point
at $(g^*_+,f^*)$,
labeled `SC' in the figure,
which governs the superconducting phase
transition.
The zero charge non-trivial fixed point labeled
`XY' governs the superfluid $^4$He transition with XY critical
exponents. This fixed point is unstable for arbitrarily small
charge. There is a second charged fixed point labeled `T' which
is infrared stable only along the line flowing from the Gaussian
fixed point to it. This fixed point is called the {\it tricritical}
fixed point and the line of infrared stability is a
{\it tricritical line}. The tricritical fixed point has coordinates
$(g^*_-,f^*)$. The tricritical line separates the left-hand region where the
phase transition is first-order
from the right-hand region where it is second-order. The tricritical fixed
point in the flow diagram is consistent with the proposed phase diagram
obtained from duality arguments, where the existence of a tricritical
point was predicted by one of us\cite{Kleinert}  in 1982,
and recently confirmed by Monte
Carlo simulations \cite{SUD}.

Let us compute the anomalous dimension for the case $\Nf{}=4$ and $\Nb{}=2$.
This is given by the fixed-point value of
the RG function

\begin{equation}
\label{gamma1}
\gamma_\phi=\mu\frac{\partial\ln Z_\phi}{\partial\mu},
\vspace{-2mm}
\end{equation}
yielding
\begin{equation}
\vspace{-2mm}
\label{etaphi}
\eta_\phi\equiv\gamma_\phi(f^*,g^*_+)=-\frac{9
\epsilon}{17}\approx-0.53\epsilon.
\end{equation}
Remakably, the anomalous dimension
$\eta_\phi$ is {\em negative\/} as in the
GL model, where it
was for a long time a great puzzle,
explained only
recently
as a consequence of
momentum space instabilities in the order field correlation
function \cite{KleinNog,Nogueira,Hove}.

In order to evaluate a second critical exponent
such as $\nu$ we need
another RG function

\begin{equation}
\label{gammam}
\gamma_m=\mu\frac{\partial}{\partial\mu}\ln\left(\frac{Z_m}
{Z_\phi}\right).
\end{equation}
At one-loop order, $\gamma_m$ is found to be

\begin{equation}
\label{gammam1loop}
\gamma_m=\frac{1}{16\pi^2}[6f-(\Nb{}+2)g].
\end{equation}
The critical exponent $\nu$ is obtained from the infrared stable
fixed point value of the function
$\nu_\phi=1/(2+\gamma_m)$
Expanded to order $\epsilon$ with $\Nb{}=2$
and, say, $\Nf{}=4$, we have $\gamma_m^*=0.024\epsilon$. 
In three dimensions, $\epsilon=1$, 
such that $\nu\approx 0.506$.

Below $T_c$, the gauge field acquires a mass whose inverse
is the penetration depth $\lambda$. The ratio between
the $\lambda$ and $\xi$ defines the Ginzburg parameter $\kappa$.
Its square can be expressed in term of the coupling constants as
$\kappa^2=g/2f$. At the mean-field level type I and type II
superconductivity are observed for $\kappa<1/\sqrt{2}$ and
$\kappa>1/\sqrt{2}$, repsectively. Fluctuations will renormalize
this
 separation point
 to $\kappa_-^{*2}\equiv g^*_-/2f^*$. The value of
$\kappa^2$ at the superconducting fixed point is given by
$\kappa_+^{*2}=g^*_+/2f^*$. For $\Nb{}=2$ and
$\Nf{}=4$, we have

\begin{equation}
\label{kappas}
\kappa^*_-=\frac{1.24}{\sqrt{2}},~~~~~~~\kappa^*_+=\frac{1.77}{\sqrt{2}}.
\vspace{-2mm}
\end{equation}
Both values are {\em above\/} the mean-field 
GL value $1/\sqrt{2}$, in contrast to
the theoretical
\cite{Kleinert} and the Monte Carlo numbers
\cite{SUD} in the GL model.

In Table I we show the values of critical exponents and the Ginzburg
parameter for $\Nb{}=2$ and $\epsilon=1$
for growing values of $\Nf{}$. We see that 
anomalous dimensions approach zero as $\Nf{}$ becomes large.
The anomalous dimension
tend to zero with increasing $\Nf{}$ much more rapidly than the
other quantities in Table I. Note that $\kappa^*_-$ decreases
with $\Nf{}$ while $\kappa^*_+$ increases with $\Nf{}$. Interestingly,
the critical exponent $\nu$ does not change very much with
$\Nf{}$, attaining quickly a limit value $\nu\approx 0.625$.

The limit $\Nf{}\to\infty$ at fixed $\Nb{}$ can be done analytically
in
the above equations:
$\lim_{\Nf{}\to\infty} f^*|_{\Nb{}=2}\to 0$, 
$\lim_{\Nf{}\to\infty} g^*_-|_{\Nb{}=2}\to 0$, 
and $\lim_{\Nf{}\to\infty} g^*_+|_{\Nb{}=2}\to 8\pi^2/5$. 
This implies that for many fermions, $\kappa^*_+\to\infty$
while $\kappa^*_-\to 0$ for $\Nb{}=2$. The limiting
critical exponents  are $\eta_\phi=0$ and
$\nu=0.625$. Since $\kappa^*_+\to\infty$ as
$\Nf{}\to\infty$ at $\Nb{}=2$ fixed, this limit is an extreme type II limit
in our model. The limiting value $\nu=0.625$ equals
 the one-loop result for the XY-model in the
$\epsilon$-expansion \cite{KS}.

\begin{table}
\caption{Critical exponents and values of the Ginzburg parameter $ \kappa $
at the tricritical and superconducting fixed point for $\epsilon=1$ and
$\Nb{}=2$ for several values of $\Nf{}$.}
\begin{ruledtabular}
\begin{tabular}{ccccc}
$\Nf{}$ &$\eta_\phi$ &$\nu$ &$\kappa^*_-(T)$ &$\kappa^*_+(SC)$\\
\hline
4& -0.53 & 0.506 & $1.24/\sqrt{2}$ & $1.77/\sqrt{2}$ \\
5& -0.43 & 0.54 & $1.09/\sqrt{2}$ & $2/\sqrt{2}$ \\
6& -0.36 & 0.56 & $1.01/\sqrt{2}$ & $2.17/\sqrt{2}$ \\
10& -0.21 & 0.59 & $0.82/\sqrt{2}$ & $2.68/\sqrt{2}$ \\
15& -0.15 & 0.6 & $0.69/\sqrt{2}$ & $3.17/\sqrt{2}$ \\
20& -0.11 & 0.61 & $0.61/\sqrt{2}$ & $3.58/\sqrt{2}$ \\
100& -0.02 & 0.623 & $0.29/\sqrt{2}$ & $7.47/\sqrt{2}$ \\
1000& -0.002 & 0.625 & $0.1/\sqrt{2}$ & $23.19/\sqrt{2}$
\end{tabular}
\end{ruledtabular}
\end{table}

Let us compare the result in $d=4-\epsilon$ dimensions to the
fixed dimension RG approach. Instead computing the $\beta$-functions
for $\epsilon$ small, we can set $m^2=0$ and compute the
Feynman integrals for any dimension $d\in(2,4)$. The $\beta$-functions
are in this case given at one loop by

\begin{equation}
\vspace{-2mm}
\vspace{-2mm}
\label{betaf2}
\beta_f=(4-d)\{-f+[8\Nf{}A(d)+\Nb{}B(d)]f^2\},
\end{equation}

\begin{eqnarray}
\vspace{-2mm}
\label{betag2}
\!\!\!\!\!\!\!\!\!\!\beta_g&=&(4-d)\bigg\{-g+C(d)\nonumber \\
&\times &\left[-2(d-1)fg+\frac{\Nb{}\!+\!8}{2}g^2+
2(d\!-\!1)f^2\right]\bigg\},
\end{eqnarray}
where

\begin{equation}
\vspace{-2mm}
A(d)=\frac{\Gamma(2-d/2)\Gamma^2(d/2)}{
(4\pi)^{d/2}\Gamma(d)},
\end{equation}

\begin{equation}
\vspace{-2mm}
B(d)=-\frac{\Gamma(1-d/2)\Gamma^2(d/2)}
{(4\pi)^{d/2}\Gamma(d)},
\end{equation}

\begin{equation}
C(d)=\frac{\Gamma(2-d/2)\Gamma^2(d/2-1)}
{(4\pi)^{d/2}\Gamma(d-2)}.
\end{equation}
The RG functions $\gamma_\phi$ and $\gamma_m$,
are at one loop:

\begin{equation}
\vspace{-5mm}
\label{gammaphid}
\gamma_\phi=(1-d)(4-d)C(d)f,
\end{equation}

\begin{equation}
\label{gammamd}
\gamma_m=(\Nb{}+2)(d-4)C(d)g/2-\gamma_\phi.
\end{equation}
Since we are working at the critical point, the RG function
$\gamma_m$ above is obtained
from an insertion
of the composite field $|\phi|^2$
 into the two-point function.

As a cross check we set $d=4-\epsilon$ and expand to first order in
$\epsilon$,
and verify that
the $\beta$-functions
 (\ref{betaf2}) and (\ref{betag2})
 reduce correctly to
the previous (\ref{betaf1}) and (\ref{betag1}), respectively.

In the absence of fermions,
the critical
value of $\Nb{}$ above which charged fixed points exist
 for $d=3$
 is
 $\Nb{}_c=103.4$, much
smaller than the value given in
the $\epsilon$-expansion, $\Nb{}_c=365.9$. On the other hand, when
we set $\Nb{}=2$ the critical number of fermions is larger than in
the $\epsilon$-expansion, being given by $\Nf{}_c=4.44$. On
the basis of the $\epsilon$-expansion result, we can give the
scheme independent estimate as $\Nf{}_c=4\pm 0.5$.

In Table II we show the values of critical exponents and
Ginzburg parameter at $d=3$ and $\Nb{}=2$ for several values
of $\Nf{}$. Qualitatively we observe the same behavior as in
Table I.

\begin{table}
\caption{Critical exponents and values of the Ginzburg parameter
at the tricritical and superconducting fixed points for $d=3$ and
$\Nb{}=2$ for several values of $\Nf{}$.}
\begin{ruledtabular}
\begin{tabular}{ccccc}
$\Nf{}$ &$\eta_\phi$ &$\nu$ &$\kappa^*_-$ &$\kappa^*_+$\\
\hline
5& -0.36 & 0.53 & $1.13/\sqrt{2}$ & $1.6/\sqrt{2}$ \\
6& -0.31 & 0.55 & $1/\sqrt{2}$ & $1.79/\sqrt{2}$ \\
10& -0.19 & 0.59 & $0.79/\sqrt{2}$ & $2.28/\sqrt{2}$ \\
15& -0.13 & 0.6 & $0.65/\sqrt{2}$ & $2.7/\sqrt{2}$ \\
20& -0.1 & 0.61 & $0.6/\sqrt{2}$ & $3.08/\sqrt{2}$ \\
100& -0.02 & 0.623 & $0.27/\sqrt{2}$ & $6.44/\sqrt{2}$ \\
1000& -0.002 & 0.625 & $0.09/\sqrt{2}$ & $20.06/\sqrt{2}$ 
\end{tabular}
\end{ruledtabular}
\end{table}

The large $\Nf{}$ limit at fixed $\Nb{}=2$ and $d=3$ gives $f^*=0$ and
$\lim_{\Nf{}\to\infty}g^*_+|_{\Nb{}=2}\to 8/5$. 
Thus, we obtain the large $\Nf{}$ value of $\nu$ at $\Nb{}=2$ and $d=3$:
$\nu=5/8\approx 0.625$.
As before, we have $\eta_\phi=0$ at $\Nf{}=\infty$.

The fermionic  Lagrangian ${\cal L}_{\rm F}$
control
 the influence of the thermal magnetic fluctuations, suppressing them
with increasing $\Nf$.
An interesting physical case in $d=3$ dimensions has fermion number
$\Nf{}=10$, where
 the critical exponents and values of
$\kappa$ at the tricritical and superconducting fixed points are
close to the predicted\cite{Kleinert}
and Monte Carlo -measured values \cite{SUD,Olsson,Nguyen}
for the pure GL
model. The critical exponent $\nu$
obtained from Monte Carlo simulations \cite{Olsson}  has the
 $XY$ model value
$\nu\simeq 0.67$, as
 predicted \cite{Kiometzis} from
 disorder field theory
of superconductors\cite{GFCM}. The anomalous dimension
$\eta_\phi$ obtained from Monte Carlo simulations is \cite{Nguyen}
$\eta_\phi\simeq -0.18$. The value of $\kappa$ at the tricritical
point predicted from the disorder field theory is
\cite{Kleinert} $\kappa^*_-=0.79/\sqrt{2}$, confirmed
by
recent
the Monte Carlo
simulations giving $\kappa^*_-=0.76/\sqrt{2}$. \cite{SUD}
The results
 in Table II,
show  amazing agreement of the theoretical values
 of $\eta_\phi$ and
$\kappa^*_-$ in the present model with $\Nf{}=10$. The critical
 exponent $\nu=0.59$ deviates slightly
from the one-loop $XY$ result. Presently we don't know if this 
result is just a lucky accident. Anyway, such a coincidence 
deserves further study.  

For $\Nf{}<4$ the model has no second-order phase transition but
interesting physical effects can be expected due to chiral
symmtery breaking.
At $d=3$ the Lagrangian (\ref{L}) has a chiral symmetry \cite{Pisarski}
$\psi\to\exp(i\gamma_{3,5}\theta)\psi$, with

\[ \gamma_3 = \left( \begin{array}{cc}
0 & I \\
I & 0 \end{array} \right),
{}~~~~~~\gamma_5 = \left( \begin{array}{cc}
0 & I \\
-I & 0 \end{array} \right) \]
where $I$ is a $2\times 2$~ unit matrix. In the absence of scalar bosons,
this symmetry is spontaneously broken for $\Nf{}<32/\pi^2\approx3.24$
and a fermion mass is dynamically generated.
\cite{Appel} According to Kim and Lee \cite{Kim}, coupling to bosons
reduce this upper bound by factor of two and one has $\Nf{}<16/\pi^2\approx
1.62$.
This lies below the critical value
$\Nf{}_c=4.44$ for the existence of charged fixed points.
The critical behavior described in this paper
 is thus apparently
not affected by the chiral symmetry breaking. However, we may wonder
if the dynamical mass generation in the chirally symmetry-broken phase
can generate new
fixed points in our system.
This point is very important concerning recent theories of 
the pseudogap state in the cuprate superconductors 
\cite{Tesanovic}.

FSN acknowledges the financial support of the Alexander von 
Humboldt foundation.

\end{document}